\documentclass[aps,showpacs,twocolumn,prl,toolkits]{revtex4}

\usepackage{mathrsfs}
\usepackage{amsmath}
\usepackage{graphicx}
\usepackage{epstopdf}
\usepackage{subfigure}

\begin{document}

\title{Heavy baryons in the Skyrme model: the role of highly anharmonic collective motion}

\author{Thomas D. Cohen}
\email{cohen@physics.umd.edu}

\affiliation{Department of Physics, University of Maryland,
College Park, MD 20742-4111}

\author{Paul M. Hohler}
\email{pmhohler@physics.umd.edu}

\affiliation{Department of Physics, University of Maryland,
College Park, MD 20742-4111}

\begin{abstract}
In the heavy quark and large $N_c$ limits, ordinary (exotic) heavy
baryons can be considered as bound states of heavy mesons
(anti-mesons)and chiral solitons. In these limits, the heavy
mesons (or anti-mesons) and the chiral solitons are extremely
heavy and are presumed to fall to the bottom of the effective
potential. Previous studies have approximated the effective
potential as harmonic about the minimum.  However, when realistic
masses for the heavy meson and chiral soliton are considered,
longer range parts of the effective potential can become relevant.
In this paper, we show that these longer-ranged  effects yield
effective wave functions which are qualitatively very different
from those expected from the combined large $N_c$ and heavy quark
limits. These potentials can support bound heavy pentaquarks under
some conditions. The consequence of these new energy states and
wave functions on the Isgur-Wise function for the semileptonic
decay of heavy baryons is also considered.
\end{abstract}


\maketitle

\section{Introduction}

It has been known for a number of years that heavy baryons can be
constructed in the heavy quark and large $N_c$ limits from the
binding of heavy mesons with light chiral solitons
\cite{jmw,glm,jm,ck1,ck2}. In these limits, both the heavy meson
and the chiral soliton have large masses, and it is legitimate to
describe the system in terms of a collective degree of freedom
between the heavy meson and baryon. The large masses drive the
particles to the bottom of the effective potential. In most cases
considered, this minimum occurs at the origin of the relative
coordinate, causing both particles to be situated on top of one
another. Previous attempts at describing heavy baryon physics
within this model assumed that particles would only experience
small harmonic perturbations in the potential away from their
minimum. The previous works showed that the channels which
constituted physical heavy baryons had attractive potentials while
those channels which did not correspond to physical particles had
repulsive potentials \cite{jmw,glm,jm}. Since the physical world
does not correspond to the extreme heavy quark and large $N_c$
limits, which guided the previous work, it is interesting to ask
the extent to which realistic masses  drive the particles to the
bottom of the potential (as expected in the combined limit) when
the harmonic approximation to the effective potential is replaced
by the leading order potential to all distances. In this paper, we
look at a class of corrections to this picture. Similar issues
have arisen for nucleon-nucleon forces in large $N_c$
\cite{boris}. We continue to assume that the heavy quark mass and
$N_c$ are large enough to enable us to describe the dynamics in
terms of a collective degree of freedom which is describable as a
non-relativistic effective potential. However, we do not
automatically assume that the masses are so large as to drive the
particles into the vicinity of the the minimum of the potential.
Instead, we focus on examining the consequences of considering the
complete potential with realistic particle masses on heavy baryon
physics. As we will show, the physics is qualitatively quite
different from what one would expect if the world were close to
the idealized limit.

Baryons described as solitons in a chiral Lagrangian were first
considered by Skyrme \cite{skyrme}. These chiral solitons have the
correct quantum numbers as baryons in the large $N_c$ limit
\cite{witten} when the Wess-Zumino term is included \cite{wz}.
While the extensive early calculations \cite{anw} focused on
$SU(2)$ solitons, the theory was also extended to include baryons
with strange quarks. This was done by either considering the
strange quark as light, and extending the $SU(2)$ chiral fields to
$SU(3)$ \cite{guadagnini,manohar}, or  by considering the strange
quarks as heavy, compared to the lighter quarks, and strange
baryons as bound states of the $SU(2)$ soltion and K-mesons
\cite{callan1,callan2}.

To extend the theory to include even heavier quarks, such as charm
and bottom, the former approach of extending the soliton group
structure is not reasonable since the mass of the heavy quarks is
vastly different from the lighter three quarks. Therefore, the
appropriate manner is the latter, put forth in the strange case by
Callan and Klebanov, where the heavy baryon states are considered
bound states of light quark chiral solitons and heavy mesons. For
simplification, we will focus our attention on models made from
$SU(2)$ chiral solitons, as was done in previous work
\cite{jmw,glm,jm}. Furthermore, the theory must also exhibit heavy
quark symmetry\cite{wise,bd,yan}.  Thereby, the heavy meson is
treated by heavy quark effective theory (HQET). HQET treats the
heavy mesons as the dynamical degrees of freedom and provides a
systematic expansion, in powers of the hadronic scale, $\Lambda$,
over the heavy quark mass, $m_Q$, to examine the Lagrangian and
subsequent interactions.

This method can also be used to bind heavy anti-mesons with chiral
solitons to form pentaquark states.  Such calculations have been
performed previously, for the case of strange quarks
\cite{klebanov}. This work showed that the pentaquark channel was
not sufficiently attractive to yield bound pentaquark states or
prominent resonances. Previously, it has been argued on general
model-independent grounds  that in the extreme heavy quark and
large $N_c$ limits, bound heavy pentaquark states must exist,
primarily because the heavy particles fall to the bottom of the
potential well \cite{hohler1}. However, it was shown that unlike
in the extreme heavy quark and large $N_c$ limits in which the
long-range one-pion-exchange potential automatically binds heavy
pentaquarks regardless of the details of the short distance
interactions, for physical masses the existence of bound
pentaquarks was highly sensitive to the details of the short
distance interaction. The Skyrme type chiral soliton models
provide one framework for treating the short distance physics, and
thereby it is interesting to determine if such theories could
support heavy pentaquark binding. More significantly, the
experience with the exotic channels suggests the possibility that
the behavior in the non-exotic channels may also be substantially
different from the limit on which the standard Skyrme analysis is
based.

The interest in determining the heavy baryon energy spectrum and
wave function within a chiral soliton model has also led to
studies of the Isgur-Wise function for transitions between heavy
baryons. Formally, the Isgur-Wise function describes the
non-perturbative physics associated with the form factor of the
semileptonic weak decay of $\Lambda_b \rightarrow \Lambda_c e^-
\bar{\nu}_e$ \cite{isgurwise}. Traditionally is is expressed as a
function of the transition velocity. If the dynamics of the heavy
baryons are known exactly, this function can be calculated for all
velocities. The Isgur-Wise function has been calculated previously
when the effective potential was approximated as harmonic both in
the context of the Skyrme model\cite{jm} and in a
model-independent context\cite{ck1,ck2}.  This paper considers the
longer range effects of the interaction between the heavy and
light degrees of freedom and leads to a non-universal form of the
Isgur-Wise function dependent on the details of the interaction.

We are by no means the first group to consider applying the Skyrme
model to attempt to explain either heavy baryons or exotic
particles. In addition to \cite{jmw,glm,jm}, heavy baryons
spectroscopy within the Skyrme model has been previous considered
by the work of Oh and his collaborators \cite{oh}. In recent
years, a variety of work associated with pentaquarks states has
been presented in the literature. These works have included heavy
pentaquarks, in the context of several different models,
\cite{hohler1,wessling1,wessling,stancu}, and light pentaquark
states either in the context of the Skyrme model,
\cite{diakonov,walliser,cohen2,cherman,trampetic1}, or in terms of
other models, \cite{jaffe,kopelovich,lipkin,cohen3}. Exoctic
dibaryon states have also been examined using the Skyrme model
\cite{balachandran}. Additionally, decays of heavy baryons have
been previously considered within the Skyrme model,
\cite{trampetic2}.

The overall goal of this paper is to explore the properties of
heavy baryons in the Skyrme model in which the collective degree
of freedom between the heavy meson and the remainder of the system
is beyond the harmonic region of the ideal heavy quark and large
$N_c$ limits. In the next section, we derive the complete
effective potential and show that when the full potential is
considered, for realistic parameters, the heavy baryon wave
function extends well beyond the region where the harmonic
approximation is applicable. This will be followed by a
demonstration that the simplest type of interactions do not create
bound states with realistic binding energies for the heavy
baryons---indeed they give the same mass for ordinary heavy
baryons and pentaquarks. This motivates the inclusion of
additional terms in the potential terms which allow for the
correct binding energies and distinguish between heavy baryon and
heavy pentaquark states. The consequences of these new potentials
will be considered. Lastly, we will examine how these newly
calculated wave functions for the heavy baryons influences the
Isgur-Wise function.

\section{Derivation of the effective potential}

The framework of this paper is based on the standard treatment of
heavy baryons in the Skyrmion in terms of a collective degree of
freedom between the heavy meson and the remainder of the system
\cite{jmw,glm,jm}.   To begin the analysis of heavy baryons, the
effective potential for this degree of freedom needs to be
determined. The relevant Lagrangian for this purpose can be
divided into the soliton sector and the HQET sector,
\begin{equation}
\mathcal{L} = \mathcal{L_{\text{Skyrme}}} +
\mathcal{L_{\text{HQET}}}.
\end{equation}
The soliton portion determines the dynamics of the ordinary
baryons.  For concreteness we will consider the simplest such
model; the one originally proposed by Skyrme:
\begin{widetext}
\begin{equation}
\mathcal{L_{\text{Skyrme}}}=\frac{1}{16} \, f_\pi^2 \, \text{Tr}
[\partial_\mu \Sigma^\dag
\partial^\mu \Sigma] + \frac{1}{32 e^2} \, \text{Tr}[(\partial_\mu \Sigma)
\Sigma^\dag,(\partial_\nu \Sigma)\Sigma^\dag]^2+\frac{1}{8} \,
m_\pi^2 f_\pi^2 \, (\text{Tr}[\Sigma]-2),
\end{equation}
\end{widetext}
where $f_\pi$ is the pion decay constant, $e$ is the Skyrme
parameter, $m_\pi$ is the pion mass, and $\Sigma$ is the chiral
soliton field \cite{skyrme}. The last term provides the pion with
a mass and fixes it to its physical value.  As we will see below,
this simplest model is not adequate phenomenologically.  However,
we will begin with an analysis of this model and consider more
sophisticated models subsequently.

Conventionally, the chiral soliton is treated classically as a
first approximation and the ansatz taken for the form of the
chiral field is,
\begin{equation}
\Sigma_0(\vec{x}) = \text{Exp}[i \vec{\tau} \cdot \hat{x} \;
F(r)],
\end{equation}
where $F(r)$ is the profile function determined by minimization of
the soliton mass.   Since this solution breaks both rotational and
isorational invariance, any rotation within the isospin space will
lead to a degenerate classical solution. That is,
\begin{equation}
\Sigma(\vec{x}) = A \, \Sigma_0 (\vec{x}) \, A^{-1}
\end{equation}
 where $A$ is an $SU(2)$ matrix of the form $A = a_0
+ i \vec{\tau} \cdot \vec{a}$ where $a_0$ and $\vec{a}$  (with
$a_0^2 + \vec{a} \cdot \vec{a}=1$) leads to a degenerate solution.
This classical degeneracy is lifted when $a_0$ and $\vec{a}$ are
promoted to quantum collective variables leading to baryons with
the physical quantum numbers\cite{anw}.

However, there is a subtlety concerning the use of the
quantization scheme of ANW for the regime considered here.  There
are two types of collective motion---the rotational/isorotational
degrees of freedom of ANW and the collective vibration of the
heavy meson off of the baryon.   Provided that the natural time
scales for these types of motion are very different they decouple.
Formally, in the large $N_c$ and heavy quark limits, when the
collective vibrational degree of freedom resides near  the bottom
of the potential well and vibrates harmonically its natural time
scale goes as $\lambda^{-1/2}$ \cite{ck3} while the time scale of
the ANW collective coordinate is $\lambda^{-1}$, where $\lambda$
is an overall counting parameter for the combined limit:
\begin{equation}
\frac{1}{N_c} \sim \frac{M_H}{\Lambda} \sim \lambda
\end{equation}
where $\Lambda$ is the characteristic hadron mass scale. Formally,
in the ideal limit, $\lambda \rightarrow 0$ and the two scales
decouple.  However, in the present study where we are explicitly
looking for effects beyond the $\lambda \rightarrow 0$ limit the
situation is more problematic.

As a practical matter, in order to proceed, we will assume that
the motions do decouple. This can be justified in part on
empirical grounds.  The splitting between the $\Sigma_c$ and the
$\Lambda_c$ in the extreme limit would be ascribed to a rotation
excitation and hence of order $\lambda$.  The splitting between
the first negative parity $\Lambda_c$ (the $\Lambda_c$(2593)) and
the $\Lambda_c$ is vibrational (in the ideal limit) and hence is
of order $\lambda^{1/2}$ which is perimetrically higher than the
rotational excitation. In practice $M_{\Sigma_c}- M_{\Lambda_c}
\approx 170$ MeV while $M_{\Lambda_c(2593)}- M_{\Lambda_c} \approx
320$ MeV. Thus, the ordering is what one expects.  However, it is
by no means obvious that one can legitimately take 320 MeV to be
considered to be qualitatively large compared to 170 MeV.

There is another reason why it is reasonable to treat the motions
as though they decouple for certain qualitative purposes.  If our
goal is to assess whether the standard Skyrmion treatment for
charmed and bottom baryons (which implicitly works in the
neighborhood of the combined limit) is self consistent, two issues
arise.  The first is whether the rotational and vibrational motion
decouple.  The second is whether the collective vibrational wave
function is well localized near the bottom of the effective
potential well. This paper is investigating the second question.
As a logical matter, if one finds that the collective wave
function does not remain well localized {\it under the assumption
that rotational and vibrational motion decouple}, then there are
important corrections to the standard treatment {\it regardless of
whether the assumption of decoupling is justified}. Therefore, by
showing that the collective wave function is not localized as in
the ideal limits, our assumptions that the standard quantization
method is applicable will be justified {\it a posteriori}.


Having argued that the standard quantization of the collective
coordinates is applicable for our purposes, we turn our attention
to the heavy meson sector of the Lagrangian. Near the heavy quark
limit, the heavy mesons' momentum is essentially carried entirely
by the heavy quark, and its properties are constrained by the
emergent heavy quark symmetry \cite{iwheavy}. Because of this,
there is an approximate degeneracy between the spin-0 and spin-1
heavy meson states. To incorporate this into the theory, HQET
combines these two fields into one overall field,
\begin{equation}
H_a = \frac{(1+v\!\!\!/)}{2} \, [P_{a \mu}^* \gamma^\mu - P_a
\gamma_5],
\end{equation}
where $v^\mu$ is the four-velocity of the heavy quark with
$v^2=1$, $P_{a \mu}^*$ is the heavy vector meson field, $P_a$ is
the heavy scalar meson field, and the index $a$ is a light quark
flavor index. The vector meson field is further constrained by
$v^\mu P_{a \mu}^* =0$. Additionally, the field, $\bar{H}_a$, can
be defined as
\begin{equation}
\bar{H}_a = \gamma^0 H_a^\dag \gamma^0 = [P_{a \mu}^{* \dag}
\gamma^\mu + P_a^\dag \gamma_5] \frac{(1+v\!\!\!/)}{2}.
\end{equation}
Using these combined fields as the relevant degrees of freedom,
the chiral Lagrangian for the heavy mesons can be written as,
\begin{widetext}
\begin{equation}
\mathcal{L_{\text{HQET}}}=-i \text{Tr}\bar{H}_a v_\mu \partial^\mu
H_a + \frac{i}{2} \text{Tr}\bar{H}_a H_b v^\mu (\Sigma^\dag
\partial_\mu \Sigma)_{ba}+ \frac{i g}{2} \text{Tr}\bar{H}_a H_b
\gamma^\mu \gamma^5 (\Sigma^\dag
\partial_\mu \Sigma)_{ba} + \ldots,
\end{equation}
\end{widetext}
where the ellipsis denotes terms with more derivatives and inverse
powers of the heavy quark mass, $1/m_H$ \cite{wise,bd,yan}. It
should be noted that in this basis of the heavy meson, the heavy
fields have an unusual transformation under parity\cite{jmw}. The
field $H_a$ transforms under parity as
\begin{equation} \label{eq:parity}
H_a(x^0,\vec{x}) \rightarrow \gamma^0 H_b(x^0, -\vec{x}) \gamma^0
\Sigma_{ba}^\dag (x^0, -\vec{x}).
\end{equation}
This transformation has the interesting property that when the
soliton and the heavy meson are located at the same point,
$\Sigma^\dag = -1$ in the transformation, while when they are
separate $\Sigma^\dag = 1$.  Thus, the heavy mesons act as though
they have negative parity at long distances (as they must) but
effectively as positive parity particles at short distances (when
the anti-quark parity is also included).

The idealized $\lambda \rightarrow 0$ limit was uniform; {\it
i.~e.}, the ordering of the large $N_c$ and heavy quark limits was
irrelevant \cite{ck3}. It is hoped that in the current problem the
ratio of the heavy quark mass to the ordinary baryon mass is
equally irrelevant provided they are both large.  If so, it is
legitimate to calculate the effective potential assuming the heavy
quark mass is infinite (but the nucleon mass is not).  We will do
that calculation here and subsequently verify that the result did
not depend on this procedure.  We will show the same effective
potential arises if the nucleon mass is taken to be infinite and
the heavy quark mass finite.

Since we are considering the complete spatially extended potential
as a function of the relative separation distance between the
particles, instead of the standard soliton ansatz, we will
consider,
\begin{equation}
\begin{split}
\Sigma(\vec{x}-\vec{x}_0(t)) &= A(t)
\Sigma_0(\vec{x}-\vec{x}_0(t)) A^{-1}(t),\\
\Sigma_0(\vec{x}-\vec{x}_0(t)) &= \text{Exp} [i \vec{\tau} \cdot
\frac{\vec{x}-\vec{x}_0}{|\vec{x}-\vec{x}_0|} \,F(|\vec{x} -
\vec{x}_0|)]
\end{split}
\end{equation}
where $\vec{x}$ is the position of the heavy meson, $\vec{x}_0$ is
the position of the soliton, and the soltion coordinate and the
collective coordinates are time dependent while the heavy meson
coordinate is not. This will fix the heavy meson to a specific
location, which we will later choose to be the origin, and allows
us to work in the rest frame of the heavy meson. Therefore the
four-velocity of the heavy meson is $v^\mu = (1,\vec{0})$.

Even with this different choice for the soliton ansatz, the only
term that produces an interaction between the heavy meson and the
chiral soliton is
\begin{equation}\label{eq:interaction}
H_I = -\frac{i g}{2} \int \!\! d^3 \vec{x} \; \text{Tr} \bar{H}_a
H_b \gamma^j \gamma^5 (\Sigma^\dag \partial_j \Sigma)_{ba}.
\end{equation}
The summation is only over the spatial coordinates in the
interaction as the spin trace with the temporal coordinate is
zero. When the appropriate soliton ansatz is inserted into the
interaction followed by some manipulation, the interaction term is
written as
\begin{widetext}
\begin{equation}
H_I = \frac{g}{2} \int \!\! d^3 \vec{x} \; \text{Tr}\bar{H}_a H_b
\gamma^j \gamma^5 \Bigl(A\biggl\{\frac{y^j}{|\vec{y}|} \hat{y}
\cdot \vec{\tau} \, \Bigl(F' - \frac{\sin(2 F)}{2|\vec{y}|}\Bigr)
+ \frac{\tau^j}{2 |\vec{y}|} \sin (2 F) + \epsilon^{jmk} y^k \tau^
m \frac{\sin^2(F)}{|\vec{y}|^2}\biggr\} A^{-1}\Bigr)_{ba},
\end{equation}
\end{widetext}
where $\vec{y}=\vec{x}-\vec{x}_0$ is the separation distance
between the soliton and the heavy meson. This can then be further
simplified by factoring out $\tau$ from each term.
\begin{widetext}
\begin{equation}\begin{split}
H_I &= \frac{g}{2} \int \!\! d^3 \vec{x} \; \text{Tr}\bar{H}_a H_b
\gamma^j \gamma^5 (A \tau^i A^{-1})_{ba} \biggl(\,\frac{y^j
y^i}{|\vec{y}|} \, \Bigl(F' - \frac{\sin(2 F)}{2 |\vec{y}|}\Bigr)
+ \frac{\sin(2 F)}{2 |\vec{y}|} \, \delta_{ij} +
\epsilon^{jik} y^k \frac{\sin^2 (F)}{|\vec{y}|^2} \biggr)\\
&= \frac{g}{4} \int \!\! d^3 \vec{x} \; \text{Tr}\bar{H}_a H_b
\gamma^j \gamma^5 \, (\tau^m)_{ba} \; \text{Tr}(A \tau^i A^{-1}
\tau^m) \, \biggl(\,\frac{y^j y^i}{|\vec{y}|^2} \, \Bigl(F' -
\frac{\sin(2 F)}{2 |\vec{y}|}\Bigr) + \frac{\sin(2 F)}{2
|\vec{y}|} \, \delta_{ij} + \epsilon^{jik} y^k \frac{\sin^2
(F)}{|\vec{y}|^2}\biggr).
\end{split}
\end{equation}
\end{widetext}
By noting that the isospin operator of the heavy meson on the H
field is,
\begin{equation} I_H^j H_a = - H_b
\frac{(\tau^j)_{ba}}{2},
\end{equation}
and the spin operator of the light degrees of freedom of the
heavy meson on the H field is,
\begin{equation} S_{lH}^k H_b =
-H_b \frac{\sigma^k}{2},
\end{equation}
along with the fact that $H_b \gamma^j \gamma^5 = -H_b \sigma^j$
in the rest frame of the $H$ field, the interaction Hamiltonian
can be written as
\begin{widetext}
\begin{equation}\label{eq:spin}
H_I= - g I_H^m S_{lH}^j \; \text{Tr} (A \tau^i A^{-1} \tau^m) \int
\!\! d^3 \vec{x} \; \text{Tr}\bar{H}_a H_a \biggl(\,\frac{y^j
y^i}{|\vec{y}|^2}\, \Bigl(F' - \frac{\sin(2 F)}{2 |\vec{y}|}\Bigr)
+ \frac{\sin(2 F)}{2 |\vec{y}|} \, \delta_{ij} + \epsilon^{jik}
y^k \frac{\sin^2 (F)}{|\vec{y}|^2}\biggr).
\end{equation}
\end{widetext}
These simplifications are similar to those performed in
\cite{glm}, whereas here we are considering the effective
potential to all distances. The integral can now be performed by
explicitly fixing the heavy meson to be located at the origin by
equating $\text{Tr} \bar{H}_a H_a$ and $-\delta(\vec{x})$. Upon
replacing the soliton position label, $\vec{x}_0$, with the more
standard $\vec{r}$, the effective potential reads:
\begin{widetext}
\begin{equation} \label{eq:bp} H_I = V(\vec{r}) = g I_H^m S_{lH}^j\;  \text{Tr}(A \tau^i A^{-1}
\tau^m) \biggl(\frac{r^j r^i}{r^2}\, \Bigl(F' - \frac{\sin(2F)}{2
r}\Bigr) + \frac{\sin(2F)}{2 r} \,\delta_{ij} - \epsilon^{jik} r^k
\frac{\sin^2(F)}{r^2}\biggr).
\end{equation}
\end{widetext}

The effective potential that we have just derived in
Eq.~(\ref{eq:bp}) has three major aspects which are interconnected
with each other. First, the potential is dependent on the spin and
isospin of the light quarks in the heavy meson.  Secondly, the
term, $\text{Tr}(A \tau^i A^{-1} \tau^m)$, is related to the spin
and isospin of the chiral soliton and is dependent on the
collective coordinate quantization as well as the states being
considered. The last part of the potential is the spatially
dependent term. This term is a function of the separation
distance, $\vec{r}$, as well as the profile function, $F(r)$. The
profile function can be derived numerically from the chiral
soliton sector of the Lagrangian. Traditionally, it is achieved by
minimizing the mass of the soliton either in the presence of a
pion mass \cite{an} or without a pion mass \cite{anw}, subject to
the constraints that $F(0) = -\pi$ and $F(\infty) = 0$.
Furthermore, if the pion wave function is expanded in powers of
$r$ and only terms of order $r^2$ are kept in the effective
potential of Eq.~(\ref{eq:bp}), one can easily show that our
potential reduces to the one considered by Jenkins, Manohar, and
Wise, \cite{jm}. Therefore in the limit that the heavy meson and
chiral soliton are close together, the effective potential is the
same as previously considered.

Let us turn our attention to the portion of the potential that is
dependent on the chiral soliton, {\it viz.} $\text{Tr}(A \tau^i
A^{-1} \tau^m)$. This term is dependent on the spin and isospin of
the soliton, yet neither spin nor isospin are guaranteed to be
valid quantum numbers of the operator. That is, in some cases the
chiral soliton term allows mixing between nucleon and delta states
within the heavy baryon system. To simplify this issue, we will
only consider the iso-scalar heavy baryons, as in the $\Lambda_H$.
This simplifies the problem because in order to create an
iso-scalar from a heavy meson and a soliton, the soliton must have
isospin-$\frac{1}{2}$. Additionally, large $N_c$ forces chiral
solitons to exhibit the property that they have the same spin and
isospin. Therefore, the only soliton that can bind with a heavy
meson to form an iso-scalar heavy baryon has spin-$\frac{1}{2}$
and isospin-$\frac{1}{2}$, or a nucleon. When the chiral soliton
is confined to the nucleon sector, it can be shown that
$\text{Tr}(A \tau^i A^{-1} \tau^m)$ is equivalent to $-8 I_N^m
S_N^i/3$, where $I_N^m$ and $S_N^i$ are the isospin and spin of
the nucleon, respectively. Making this replacement in the
effective potential and summing over repeated indices leads to a
potential operator that reads:
\begin{widetext}
\begin{equation}
V(\vec{r}) = -\frac{8 g}{3} (\vec{I_H} \cdot \vec{I_N})
\Bigl((\vec{S_{lH}} \cdot \hat{r})(\vec{S_N} \cdot \hat{r}) (F'
-\frac{\sin (2F)}{2 r}) + (\vec{S_{lH}} \cdot \vec{S_N})
\frac{\sin (2F)}{2 r} - (\vec{S_{lH}} \times \vec{S_N}) \cdot
\hat{r} \, \frac{\sin^2(F)}{r}\Bigr).
\end{equation}
\end{widetext}

Having derived a potential operator, the problem reduces to
finding the eigenvalues and eigenstates of this operator. In order
to determine the eigenstates of this potential operator, let us
consider states labelled by the total isospin, $I$, the total
spin, $s$, and the spin of the light degrees of freedom, $s_l$.
These states can be written as $|I,s,s_l\rangle$. For a total
isospin-0, we can construct three states;
$|0,\frac{1}{2},0\rangle$, $|0,\frac{1}{2},1\rangle$, and
$|0,\frac{3}{2},1\rangle$.  From the potential, it is clear that
total isospin is a good quantum number for the states, however,
the spin of the light degrees of freedom is not obviously a good
choice here as the cross product term changes the spin state.
Therefore, instead of the simple state $|I,s,s_l\rangle$, the
appropriate wave function that should be considered has the form:
\begin{equation} \label{eq:spin0}
\Psi_{\Lambda 0} = [1 + D(r) \vec{r} \cdot (\vec{S_{lH}} \times
\vec{S_N})] |0,\frac{1}{2},0\rangle \phi(\vec{r})
\end{equation}
for the $|0,\frac{1}{2},0\rangle$ state, and
\begin{widetext}
\begin{equation} \label{eq:spin1}
\Psi_{\Lambda 1} = [1 + D'(r) \vec{r} \cdot (\vec{S_{lH}} \times
\vec{S_N}) + E'(r) (\vec{S_{lH}} \cdot \vec{r}) (\vec{S_N} \cdot
\vec{r})] |0,\frac{1}{2},1\rangle \phi(\vec{r})
\end{equation}
\end{widetext}
for the $|0,\frac{1}{2},1\rangle$ and $|0,\frac{3}{2},1\rangle$
states.  There is a degeneracy for the light quark spin-1 states
because of the degeneracy between the pseudo-scalar and vector
heavy mesons in the heavy quark limit. It can be shown that the
wave function for the light quark spin-0 state is in fact the
eigenfunction of the potential operator when
\begin{equation}\label{eq:d1}
D(r) = \frac{-2 (\cos(F)+1)}{r \sin(F)}
\end{equation}
with an eigenvalue of
\begin{equation}\label{eq:pot1}
V_{\Lambda 0}(r) = -\frac{g}{2} F'(r)+ g \frac{\sin(F)}{r}.
\end{equation}
The effective potential for the isospin-0 light quark spin-1
channel can be obtained in a similar manner.  Here, the form given
above is the eigenfunction when
\begin{equation}\label{eq:d2}
D'(r) = \frac{-4 (1+\cos(F))}{r \sin(F)} \quad \text{and} \quad
E'(r) = -\frac{4}{r^2}
\end{equation}
with the eigenvalue
\begin{equation}\label{eq:pot2}
V_{\Lambda 1}(r) =  -\frac{g}{2} F'(r) - g\frac{\sin(F)}{r}.
\end{equation}

The previous discussion was based on taking the heavy meson mass
to be arbitrarily large so that collective dynamics involved the
soliton moving.  We have argued that the resulting dynamics ought
to be independent of this assumption. To demonstrate this, the
effective potential with the soliton's position held fixed can be
calculated with the methods described above with a few caveats.
First, since the heavy meson is now moving, the kinetic energy
term of the heavy meson needs to be explicitly considered. From
HQET \cite{wisemanohar} the additional term in the Lagrangian for
the heavy meson fields is
\begin{equation}
\mathcal{L_{\text{kinetic}}} = \text{Tr} \bar{H}_a
\frac{(D_{\perp}^2)_{ba}}{2 M_H} H_b,
\end{equation}
where $D_\perp^\mu$ is the covariant derivative perpendicular to
the velocity and is defined as $(D_{\perp}^{\mu})_{ba} =
(D^\mu)_{ba} - v^\mu (v \cdot D)_{ba}$. The velocity of the heavy
meson is given by $v$, and $D^\mu$ is the covariant derivative
defined by $(D^\mu)_{ba} = \partial^\mu \delta_{ba} - \frac{1}{2}
(\Sigma^\dag \partial^\mu \Sigma)_{ba}$. The Roman indices are the
light quark flavor indices, as before. Second, even though we are
allowing the heavy meson to move, we would still want to be close
to the heavy quark limit, therefore, the heavy meson's velocity
will be small; $v^\mu = (1,\vec{\epsilon})$.

The effective potential can still be derived from the interaction
term as written in Eq.~(\ref{eq:interaction}) except the integral
is now over $x_0$--- \!\!the soliton's position rather than the
heavy meson's position. The use of only the spatial directions in
this equation is still justified since corrections to this are of
order $\epsilon$, which will remain small. The calculation
proceeds as previously illustrated until Eq.~(\ref{eq:spin}). The
substitution of the spin of the light quark in the heavy meson
from the previous formula is still possible with again corrections
of $O(\epsilon)$. At this point, the integral can be performed
analogously as before, but by fixing the soliton's position to be
$x_0 = 0$. However, unlike before, we are left with the term
$\text{Tr} \bar{H} H$ in the expression of the effective
potential.  This term is dependent on the heavy meson wave
function.  However, the heavy meson wave function can be expressed
as an exponential, {\it i.e.}, $H \sim \text{Exp} [i f(x)]$, where
$f(x)$ is some unknown function of the heavy meson's position.
With the appropriate normalization, we can thereby set $\text{Tr}
\bar{H}H = -1$. Thus we have (practically) derived the same
effective potential as in Eq.~(\ref{eq:bp}).  The careful observer
will notice that the potential derived with the soliton held fixed
is identical with Eq.~(\ref{eq:bp}) except for the sign of the
last term which we will show will lead to the same physical
system.

The eigenfunctions and eigenvalues of the new potential operator
can be constructed just as before. However, when we use the
effective potential having held the soliton fixed instead, $D(r)$
and $D'(r)$ in Eqs.~(\ref{eq:d1}),(\ref{eq:d2}) have the opposite
sign. This sign difference compensates exactly the sign difference
between the potentials discussed above. Thereby both methods lead
to the same physical effective potentials in
Eqs.~(\ref{eq:pot1}),(\ref{eq:pot2}). Since both methods lead to
the same physical effective potentials we have demonstrated the
commutativity of the large $N_c$ and heavy quark limits in this
problem.

To complete the discussion of these wave functions, we need to
establish the correspondence between the $|I,s,s_l\rangle$ states
and the physical $\Lambda_H$ states. From the states' quantum
numbers it is clear that the light quark spin-0 state,
$|0,\frac{1}{2},0\rangle$, corresponds to the ground state of
$\Lambda_H$, while the light quark spin-1 states are spin
excitations of this ground state which have yet to be observed.
The observed excited states to $\Lambda_c$, $\Lambda_c^+(2593)$
and $\Lambda_c^+(2625)$, would constitute radial excitations of
the light quark spin-0 state. However, even with these apparently
clear assignments, the parity of the heavy baryon states is not
immediately clear in this language. The wave functions for both
the light quark spin-0 and spin-1 states written above do not
appear to have definite parity. Thus when the orbital momentum
between the heavy meson and the soliton is considered in the $l=0$
state, the ground state wave function contains parts which are
characteristic of both s- and p-wave states. The states achieve
definite parity when we recall that the heavy meson state itself
is negative under parity when near the soliton and positive when
far apart, as pointed out in Eq.~(\ref{eq:parity}) and the
sentences following that equation. The wave functions in
Eq.~(\ref{eq:spin0}) and Eq.~(\ref{eq:spin1}) show that when the
heavy meson and soliton are close together, the state has a
positive parity (positive from the s-wave, positive from the heavy
meson) while when they are far apart, the state still has positive
parity (negative from the p-wave, negative from the heavy meson).
Thereby, the ground states, and thus subsequent excitation, have
the same parity as their physical particle states, which completes
the assignment between wave functions and physical states.  Note
again, the subtlety associated with orbital momentum states since
the orbital momentum is not a good quantum number. Henceforth in
this paper we will label these states by the $l$ used in the
Schr\"{o}dinger equation. This can be thought of as the orbital
momentum state when the heavy meson and soliton are close
together. In previous studies, since they were only concerned with
small motion of the potential away from the heavy meson and
soliton sitting at the same place, these long distance effects on
the parity of the states did not matter.  Thus, in their work the
states could be clearly labelled by the orbital momentum between
the heavy meson and the soliton.

We have derived the effective potential of a heavy meson and a
nucleon in terms of the profile function, $F(r)$, for the
isospin-0 light quark spin-0 and spin-1 channels. Note that these
potentials are completely radial. These potentials include short
and long distance behavior for the binding which inherently has
not been considered before. It should be noted that when both of
these potentials are examined at short distances, they reduce to
the potentials and values at the origin that have been previously
identified \cite{jmw,glm,jm}.

\section{Determination of bound states}

At this point, the effective potentials that we have constructed
can be used in a Schr\"{o}dinger equation, and the bound states
can be calculated. At the time of the previous studies the heavy
meson-soliton coupling, $g$, was undetermined.  In recent years
this has been measured to be $\approx 0.59$ from the decays of
$D^*$ meson into $D$ mesons and pion emissions \cite{cleo}. We
have assumed that this coupling is the same for B mesons (as it
should be in the heavy quark limit) since an experimental
determination via pion emission is not energetically possible. The
physical mass of the spin-0 heavy meson was used (1864 MeV for $D$
meson and 5279 MeV for $B$ meson \cite{pdg}), while the mass of
the soliton was calculated from the profile function.

\begin{figure}[tb]
\includegraphics[scale=.75]{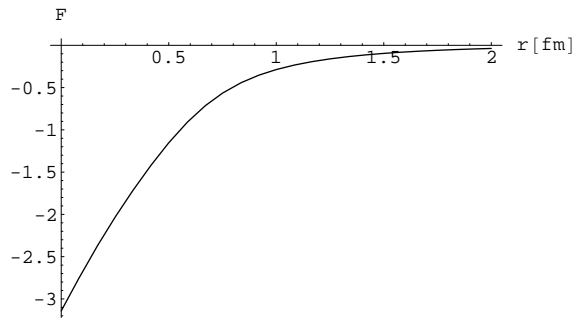}
  \caption{\label{fig:pion} Calculated Skyrmion profile function, $F(r)$.}
\end{figure}

The short and long distance structure of the profile function,
$F(r)$, was obtained by examining the differential equation that
minimized the mass of the chiral soliton. From there, the profile
function was constructed by parameterizing the functional form
consistent with the short and long distance behavior with two
parameters. These two parameters where determined by an iterative
method that minimized the mass of the soliton while keeping the
nucleon-pion coupling, $g_A$, and the pion decay constant,
$f_\pi$, constant. This procedure constructed the Skyrmion profile
function that is plotted in Fig.~\ref{fig:pion} and fixed the
Skyrme parameter, $e=4.10$, and the soliton mass, $M = 949$ MeV.

\begin{figure}[htb]
\includegraphics[scale=.8]{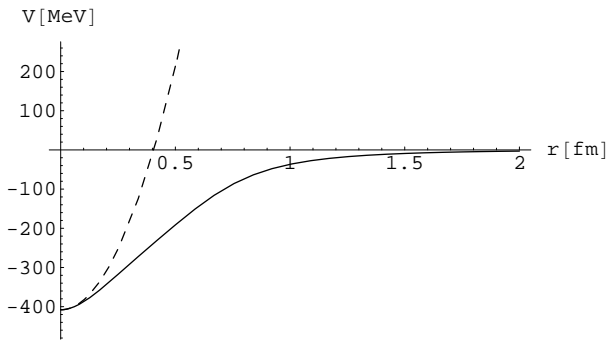}
\caption{\label{fig:pot} Light quark spin-0 state effective
potential (solid curve) with harmonic approximation (dashed curve)
as a function of separation distance.}
\end{figure}

\begin{figure}[htb]
\centering \subfigure[Charm Wave Function]{\label{fig:cwave}
\includegraphics[scale=.8]{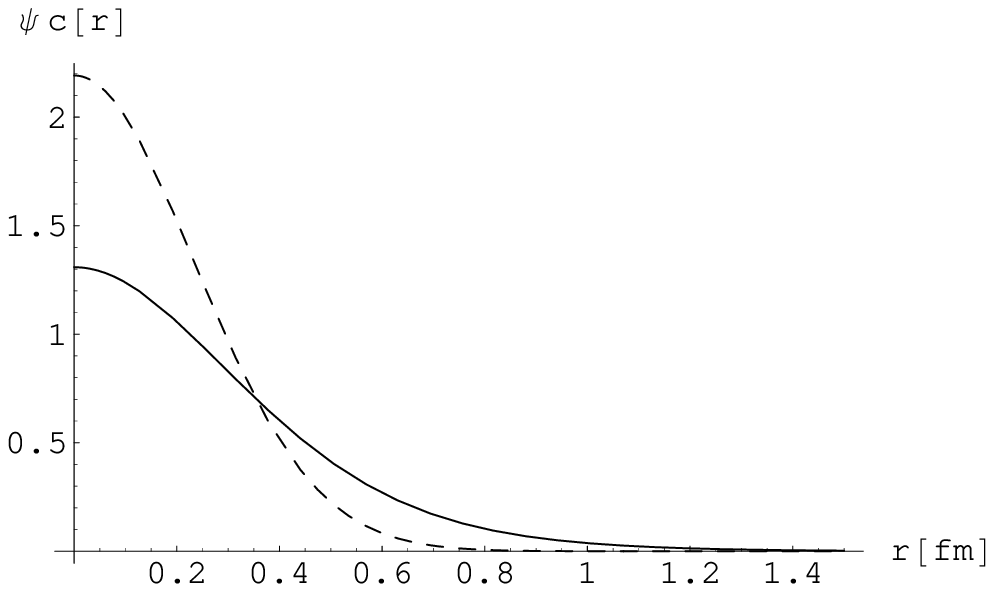}}
\subfigure[Bottom Wave Function]{\label{fig:bwave}
\includegraphics[scale=.8]{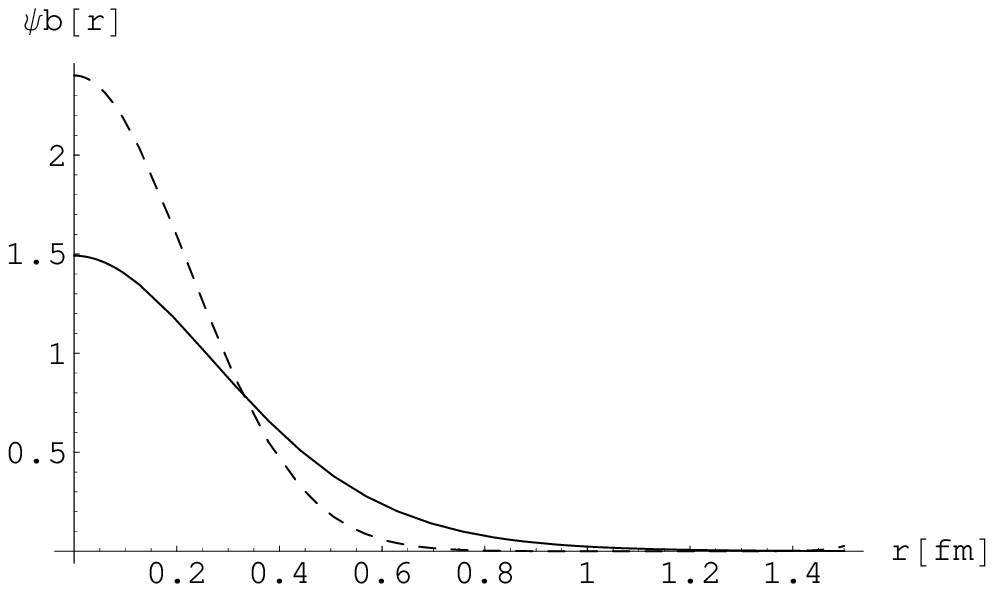}}
\caption{\label{fig:wave} (a) Calculated normalized wave function
for the ground state $\Lambda_c$ from the complete potential
(solid curve) and the harmonic approximation (dashed curve). (b)
Calculated normalized wave function for the ground state
$\Lambda_b$ from the complete potential (solid curve) and the
harmonic approximation (dashed curve).}
\end{figure}

The Sch\"{o}dinger equation with the appropriate effective
potential for this system was solved to observe the presence of
bound states. Figure \ref{fig:pot} shows the potential for the
light quark spin-0 state with the harmonic oscillator
approximation overlaid. When the equation was solved, we found for
the charm case, a binding energy of 155 MeV and for the bottom
case, a binding energy of 177 MeV. In both cases a weakly bound
radial excited state was also observed; 6.18 MeV for charm and
19.32 MeV for bottom. The observed ground states are more tightly
bound than the ground state in the harmonic oscillator
approximation. Therefore the inclusion of the entire potential
increases the binding energy and favors a stronger bound state.
Furthermore, the wave function of the nucleon in the ground state
is much broader with the extended potential compared with the wave
function of the harmonic oscillator (see Fig.~\ref{fig:wave}).
This increase in the wave function breadth indicates that the
nucleon is influenced by the long distance part of the potential.

It is clear that for this particular model, both the shape of the
wave function and the binding energy are vastly different when the
entire effective potential is considered as compared to when the
potential is assumed to be harmonic with the collective degree of
freedom tightly localized.  Large amplitude motion clearly occurs
and the standard analysis appears to be invalid for this system.

For this system, orbital excited states can also be seen. Both the
charm and bottom cases have an $l=1$ excited state; the binding
energy is 42.0 MeV  (charm) and 68.0 MeV (bottom). Neither system
appears to have a bound $l=2$ state.

For the case when the light quark system carries spin 1, no bound
states were found. The potential in this channel has a strong
repulsive core with a very shallow attractive region which appears
to be too weak to support bound states.

The previous calculations were performed using the assumption
(valid in the heavy quark limit) that vector and pseudoscalar
heavy mesons are degenerate.  This is obviously not true in the
physical case, that is, the $D$ and the $D^*$ or the $B$ and the
$B^*$ have different masses.  The calculation can be extended to
include the physical mass splittings between the heavy meson
states. When these splittings are included, there are still bound
$\Lambda_H$ states, however, the binding is weaker. The binding
energy of the ground state is reduced by 88.7 MeV for charm and
35.9 MeV for bottom.

To reiterate, these results clearly illustrate that the effective
potential is not strong enough to localize the collective variable
in the harmonic region for these models. The underlying assumption
that the system is sufficiently close to the large $N_c$ and heavy
quark limits to use the harmonic approximation is not justified.
Clearly it is important to see whether the breakdown of the
combined heavy quark and large $N_c$ limit is generic for
realistic nucleon and heavy quark masses. This is particularly
true since the model considered has serious phenomenological
flaws, as will be discussed below.  The key question is whether
the standard treatment works for ``realistic'' models.

\section{Towards the effective potential in realistic models}

The model considered in the previous section is unsatisfactory in
terms of phenomenology.  In the first place the mass of the heavy
baryons is well off from the empirical ones. The relevant issue is
not how large the fractional error is for the mass since a large
fraction of the mass is simply from the heavy quark itself.  The
relevant issue is the fractional error in the binding
energy---{\it i.e.}, the difference between the mass of a nucleon
plus a heavy meson from the mass of the heavy baryon mass.  In the
model considered above $M_n + M_D - M_{\Lambda_c}$ = 155 MeV.  In
nature it is approximately 520 MeV. This is certainly not very
good.

Moreover, the model considered above has the feature that the
interaction is identical under the exchange of a heavy meson to a
heavy anti-meson.  That is, our extended potential will give the
same bound states for ordinary heavy baryons as for heavy
pentaquark states.  Clearly this is unphysical. While heavy
pentaquark states are known to exist in the extreme heavy quark
and large $N_c$ limits\cite{hohler1}, there exists no symmetry of
QCD in that limit which implies a degeneracy between pentaquarks
and ordinary heavy baryons. Moreover, strong-interaction-stable
heavy pentaquarks have not been detected despite intensive
searches, suggesting that for realistic masses they do not exist.

If we wish to make a more realistic model it is necessary to
include additional interaction terms between the heavy meson and
the soliton which split the ordinary heavy baryon from the
pentaquark. Such an interaction should be strong enough to give
heavy baryons with approximately the correct mass. Heretofore, we
have only included an interaction term between the pion fields and
the heavy quark which was lowest order in the chiral expansion.
However, in the soliton there is no chiral power counting and thus
no necessity to restrict the interaction to this term.  The spirit
of model building in Skyrme type models is to include a small
number of terms to make the problem tractable. Although there is
no systematic power counting, the hope is that one can get
qualitatively sensible results by choosing coefficients for these
which compromise between the various observables.  The simplest
model realistic enough to get the binding of the heavy baryons
correct while pushing up the mass of the heavy pentaquarks above
threshold will require one new interaction term.

The simplest term that we can consider, which distinguishes
between interactions between heavy mesons and anti-mesons, is
coupling the light quark baryon current to the heavy quark vector
current. Note that if a heavy anti-meson binds with a nucleon, the
heavy quark current will switch sign compared with the heavy meson
case while the baryon current (associated with the nucleon) will
not. Therefore we add the term
\begin{equation}
\mathcal{L}_{\text{baryon}} = g' \text{Tr} \bar{H}_a H_a
\gamma_\mu B^\mu
\end{equation}
to the Lagrangian, with the baryon current, $B^\mu$, given by the
standard form,
\begin{equation}
B^\mu = \frac{\epsilon^{\mu \nu \alpha \beta}}{24 \pi^2} \text{Tr}
[(\Sigma^\dag \partial_\nu \Sigma)(\Sigma^\dag \partial_\alpha
\Sigma)(\Sigma^\dag \partial_\beta \Sigma)],
\end{equation}
where the notation $\epsilon_{0123}=-\epsilon^{0123}=1$ was used.
By inserting the functional form for the solitons into this term,
and working in the rest frame of the heavy (anti)meson, the
interacting potential can be derived as :
\begin{equation}
V_{\text{baryon}}(\vec{r}) = g' B^0(r) = \frac{g'}{2 \pi^2}
\frac{\sin^2(F)}{r^2} F'(r).
\end{equation}
The coupling constant, $g'$, and its relative sign are unknown,
but the sign will be chosen such that this potential is attractive
for heavy meson-nucleon interactions while repulsive for heavy
anti-meson-nucleon interactions.  In the spirit of this class of
model, we will tune $g'$ in order to get a reasonable mass for the
heavy baryon (either $\Lambda_b$ or $\Lambda_c$). Of course, this
procedure is quite {\it ad hoc}, but Skyrme type models always
require some {\it ad hoc} procedure.  One useful check on whether
the model obtained is sensible is whether the the coupling
obtained has a natural size, {\it i.e.}, whether it is of $O(1)$.

When the Sch\"{o}dinger equation with the combined potential was
used to calculate bound states, the coupling constants needed to
have bound states with physical binding energies were determined
to be -3.27 for $\Lambda_c$ and -3.34 for $\Lambda_b$. These are
natural in size. Furthermore,  the two coupling constants obtained
via fitting for the two types of heavy baryons are quite close to
one another; they differ by only 2\% suggesting that the procedure
is robust for physical values. For simplicity, the heavy meson
mass splittings were not included in this calculation.

Because of the lack of the heavy meson mass splitting, we would
expect that the first excited state to be a superposition of the
two physically observed excited states, $\Lambda_c^+(2593)$ and
$\Lambda_c^+(2625)$. It is not unbelievable to surmise that this
combined state would weight the two states by the relative spins
of the states. This would lead to a mass of the combined state of
2614 MeV or an excitation energy of 328 MeV above the ground
state. With the coupling constants considered here, the excitation
energy from the combined potential was found to be 324 MeV and 316
MeV for the charm and bottom cases, respectively. This is in
surprisingly good agreement with the expected excitation from the
physically seen states.

With the inclusion of the baryon current term to the potential,
the light quark spin-1 state becomes bound for the coupling
constants considered here. The binding energies are weak,; 125 MeV
for charm and 173 MeV for bottom, compared with their spin-0
counterparts, $521.6$ MeV and 593.4 MeV for the $\Lambda_c$ and
$\Lambda_b$, respectively. This binding is driven solely by the
size of the baryon current coupling. The presence of a bound state
in the spin-1 channel does not invalidate nor should it limit our
method. It is not unreasonable to believe that the strength of the
additional potentials that is needed to be included to drive the
binding energy down to physical values might also bind light quark
spin-1 states. Therefore the lack of any observation of these
states could be associated with the difficulty in detection and
not in the lack of the presence of these states.

A harmonic approximation can also be made for the combined
potential with the provided coupling constants. The energy levels
for the approximation are again different from that of the
combined potential and the ground state is bound weaker. An
examination of the shape of the wave function for the combined
potential compared with the harmonic wave functions shows again
that the wave functions for the combined potential are broader
than the harmonic wave functions and extend beyond the harmonic
region. We again find ourselves not driven to the effective
potential minimum by the realistic masses suggesting that the
physical states are not close to the extreme large $N_c$ and heavy
quark limits. These observations confirm the results from the
previous section that the long distance potential greatly
influences heavy baryon formation.

Unlike with our original potential, the new combined potential is
sensitive to the difference between regular heavy baryons and
heavy pentaquarks. We can examine whether a bound pentaquark state
is possible within the combined potential by flipping the sign of
the coupling constant of the baryon current term. When this is
performed, a very weakly bound pentaquark state can be found with
a binding energy of $33.4$ MeV for a charm type pentaquark, and
$42.9$ MeV for a bottom type pentaquark.  These are rather small
binding energies compared with the binding energies of the regular
heavy baryons, and are small compared to the deeply bound
pentaquark states suggested by the large $N_c$ and heavy quark
limits \cite{hohler1}. Furthermore, the potential for the heavy
pentaquark is repulsive at short distance creating the attractive
region and localization of the wave function away from the origin.
Therefore, even in the ground state, the pentaquark formed here
has a fixed separation between the heavy anti-meson and the
nucleon. However, due to the strong repulsion at short distance
this is a very delicate state. If we turn off the potential which
is identical for the heavy baryon and heavy pentaquark states, and
tune the baryon current coupling such that heavy baryons are still
bound with the appropriate binding energy, we find that the heavy
pentaquark effective potential is completely repulsive and thereby
no heavy pentaquark state is bound. The coupling constants needed
to achieve this condition were -5.52 for the charm case and -5.49
for the bottom case.  These couplings are again neither
unreasonably large nor small. This indicates that this model is
capable of binding a heavy pentaquark state, but with minor
changes it is just as reasonable not to support a bound heavy
pentaquark state.  The relative ease for the state to become
unbound furthers the point in \cite{hohler1} that heavy pentaquark
binding with physically reasonable parameters is subject to the
dynamical details of the model.

We have justified the inclusion of an additional interaction term
based upon the coupling of the heavy quark current to the baryon
current.  This interaction was derived for all separation
distances.  The coupling of the interaction was determined by
having the ground state binding energy correspond to the physical
value. With this potential in place, our simple model gives
phenomenologically reasonable results concerning ordinary heavy
baryons while simultaneously being reasonably consistent and
inconsistent with a bound heavy pentaquark state.

\section{Calculation of the Isgur-Wise Function}
The Isgur-Wise function is a universal function in the heavy quark
limit that describes the semileptonic decay of all heavy hadrons
in an SU(2$N_H$) multiplet (where $N_H$ is the number of heavy
flavors).  In our case we can consider heavy hadrons of the
$\Lambda$ type \cite{isgurwise}.  Thus the process in question is
 $\Lambda_b
\rightarrow \Lambda_c e^- \bar{\nu}_e$ \. Previous work has shown
that the Isgur-Wise function is completely determined in the
combined large $N_c$ and heavy quark limit \cite{jm}.  In this
limit the effective potential is purely harmonic in nature, and
the Isgur-Wise function would be dependent on only one parameter,
the harmonic ``spring constant'' \cite{ck1,ck2}. Therefore it has
been pointed out that if the excitation energy of heavy baryons
were measured, the spring constant would be fixed and thus
Isgur-Wise function would be completely  determined up to higher
order corrections.

Unfortunately, we have shown thus far that for realistic
parameters, the system does {\it not} remain in the harmonic
region and the expansion appears to break down. Of course, from
the wave functions we have already calculated, the Isgur-Wise
function can also be calculated, and again we would expect
distinct results from the harmonic oscillator case.

The most interesting aspect of the Isgur-Wise function near the
combined limit is related to the fact that this entire universal
function is dependent on one measurable parameter. We wish to test
the reliability of this result when one deviates from the ideal
limit, {\it viz.}, for the real world. Ideally one could test this
by using the empirical value of the excitation energy of the first
excited state plus the assumption that one is near the combined
large $N_c$ and heavy quark limits to compute the Isgur-Wise
function; this ought then to be compared with the experimental
Isgur-Wise function for an empirical test. However, the Isgur-Wise
form factors have not been measured.

In the absence of such data, we can still get some idea of how
robust is the prediction by using  ``realistic'' models.  Since
these models indicate that that system is quite anharmonic, we do
not expect a close agreement of the Isgur-Wise function with the
harmonic approximation based on the energy of the first excited
state. Instead of considering the entire Isgur-Wise function, we
will focus our attention on the curvature at zero recoil, $\rho$,
of the Isgur-Wise function.  We chose to do this as $\rho$ is
proportional to the mean radius squared in the heavy quark limit,
and thereby provides a single number with which to compare and
does so in a physically transparent way.

The Isgur-Wise function in momentum space from \cite{jm} is
\begin{equation}
\eta_0 = \int \!\! d^3 \vec{p}\; \phi^*_c(\vec{p}+m_N \vec{v}') \,
\phi_b(\vec{p}).
\end{equation}
In the case of harmonic wave functions this reduces to
\begin{widetext}
\begin{equation}
\eta_{HO}(z) = \frac{2 \sqrt{2} \mu_b^{3/8}
\mu_c^{3/8}}{(\sqrt{\mu_b}+\sqrt{\mu_c})^{3/2}}\, \text{Exp}
\Bigl[\frac{-z^2}{2 \sqrt{\kappa}}\Bigr], \quad \text{where} \quad
z\equiv \frac{m_N |\vec{v}|}{(\sqrt{\mu_b}+\sqrt{\mu_c})^{1/2}},
\end{equation}
\end{widetext}
$m_N$ is the mass of the nucleon, $\kappa$ is the harmonic
coupling, $\vec{v}$ is the transfer velocity, and $\mu_c$ and
$\mu_b$ are the reduced mass of the heavy meson-nucleon system for
either charm or bottom, respectively. We can also express the
Isgur-Wise function in terms of position space wave function as
\begin{equation}
\eta_0 = \int \!\! d^3 \vec{x} \;\psi_c^* (\vec{x})
\psi_b(\vec{x})\, \frac{\sin(\frac{m_N}{|\vec{v}||\vec{x}|})}{m_N
|\vec{x}||\vec{v}|}.
\end{equation}
These expressions for the Isgur-Wise function lead to the
following leading order expressions for the curvature at zero
recoil,
\begin{equation} \label{eq:rhoho}
\rho_{HO} \equiv \frac{\partial^2 \eta_{HO}}{\partial z^2} (z=0)
= - \frac{2 \sqrt{2} \mu_b^{3/8} \mu_c^{3/8}}{\sqrt{\kappa}\,
(\sqrt{\mu_b}+\sqrt{\mu_c})^{3/2}}
\end{equation}
for the harmonic case, and
\begin{equation} \label{eq:rhonoho}
\rho \equiv \frac{\partial^2 \eta}{\partial z^2} (z=0) =
-\frac{\sqrt{\mu_b}+\sqrt{\mu_c}}{3} \int \!\! d^3 \vec{x} \;
\psi^*_c (\vec{x}) \psi_b (\vec{x}) |\vec{x}|^2
\end{equation}
for the spatial wave function case.

We will calculate $\rho$ in three different manners.  First,
$\rho$ can be calculated by approximating the potential by a
harmonic oscillator to generate a harmonic coupling, $\kappa$.
Equation (\ref{eq:rhoho}) can then be used for this value of
$\kappa$, along with physical values of the mass variables, to
calculate $\rho$. Second, the wave functions of the ground state
of the complete potential can be used directly in
Eq.~(\ref{eq:rhonoho}) to calculate $\rho$. Of course, this is the
proper way to calculate $\rho$ and the Isgur-Wise function since
the complete potential is being considered. We would expect a vast
difference between the first two methods since we have been
observing differences in the wave function between the harmonic
wave functions and the complete potential wave functions. Lastly,
even though we know that the energy levels calculated by using the
entire effective potential are not harmonic in nature, we could
use the excitation energy of the first excited state to calculate
$\kappa$ as though it were derived from a harmonic oscillator, and
then use this value of $\kappa$ to calculate $\rho$ using
Eq.~(\ref{eq:rhoho}). That is, instead of deriving the harmonic
coupling from an approximation to the actual potential, we are
constructing a new harmonic potential, which is unrelated to the
original harmonic potential and to the complete potential, except
that the energy gap between the ground state and the first excited
state of this new potential and the complete potential are
identical.  Of course, the third method has no underlying
theoretical basis unless the system is harmonic (in which case it
will agree with the first two).  It is useful to consider it,
however, because it simulates theoretically the empirical
procedure outlined above.

\begin{figure}[htb]
\centering \subfigure[Charm Wave Functions] {\label{fig:cwavep}
    \includegraphics[scale=.8]{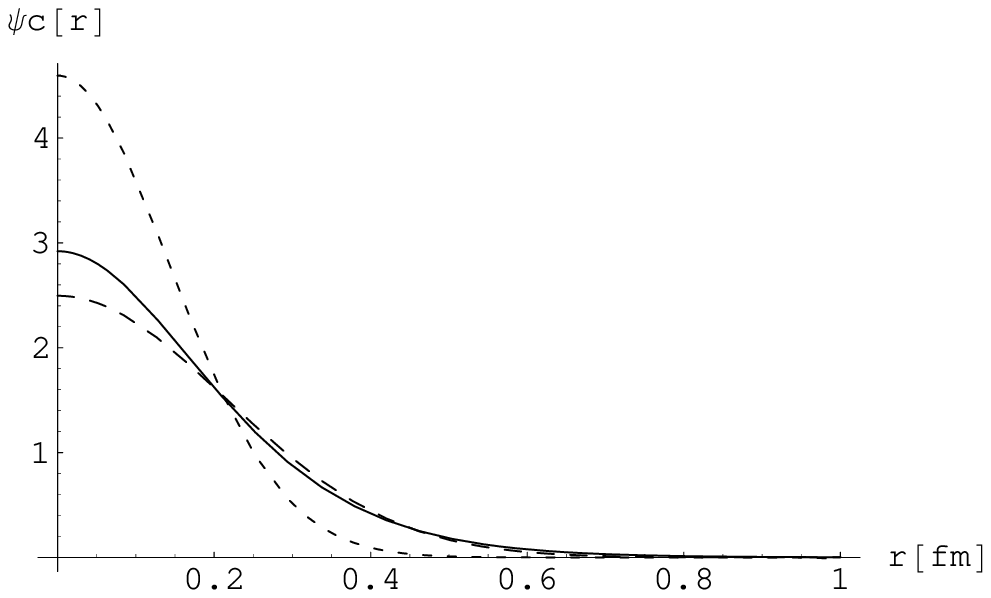}}
\subfigure[Bottom Wave Function]{\label{fig:bwavep}
    \includegraphics[scale=.8]{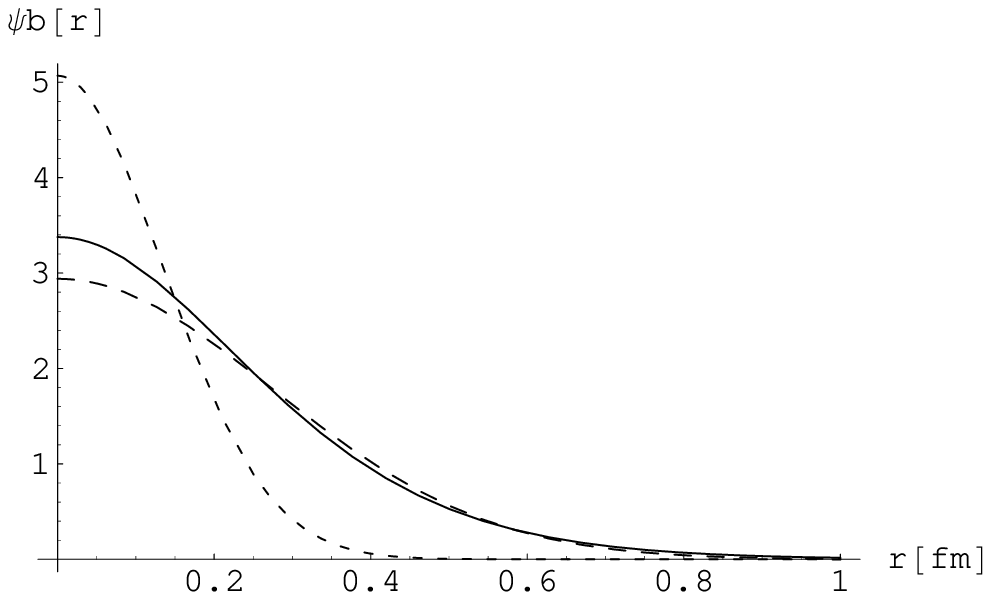}}
\caption{\label{fig:wavep} (a) Calculated normalized wave
functions for the ground state $\Lambda_c$ from the complete
potential (solid curve), the harmonic approximation (smaller
dashed curve), and the harmonic approximation from the excitation
energy (larger dashed curve). (b) Calculated normalized wave
functions for the ground state $\Lambda_b$ form the complete
potential (solid curve), the harmonic approximation (smaller
dashed curve), and the harmonic approximation from the excitation
energy (larger dashed curve).}
\end{figure}

When the curvature of the Isgur-Wise function at zero recoil,
$\rho$, was calculated in these manners, using physical mass of
the nucleon and the heavy baryons, the following results were
obtained.  The normalized wave functions are shown in
Fig.~\ref{fig:wavep}, remembering that $\rho$ is proportional to
the expectation value of $r^2$.
Method 1 yielded $-53.8 \times 10^{-6} \text{MeV}^{-3/2}$ when
$g'$ was chosen to have the appropriate coupling constant for the
bottom sector.  Method 2 yields $-121 \times 10^{-6}
\text{MeV}^{-3/2}$. Note that by using the complete wave
functions, the curvature is a factor of 2 larger than the case of
using harmonic oscillator wave functions. Lastly, Method 3, of
using the excitation energy from the full potential and assuming
it came from a harmonic oscillator potential, yields
$-111\times10^{-6} \text{MeV}^{-3/2}$ with the bottom case
coupling constant. This result is different from the Method 1 as
one might have expected, but is quite similar to the Method 2
(which gives the ``correct'' result).  This is quite surprising
since Method 3 can only be justified via the harmonic
approximation which appears to be badly violated (as seen from the
result of Method 1).  Given the fact that Method 3 simulates the
natural way to test the harmonic approximation empirically, it is
important to test whether the success of Method 3 in getting close
to the correct result is a mere numerical accident for this model
or whether it is a robust feature.  As we will see it is quite
robust.

\begin{figure}[htb]
\includegraphics[scale=.75]{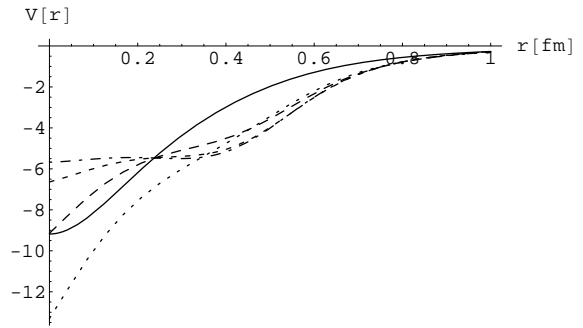}
  \caption{\label{fig:effpot} Variety of effective potentials: (a) solid
  curve; (b) long dashed curve; (c)
  short dashed curve; (d) dotted curve; (e)
  dot-dashed curved.}
\end{figure}

In order to test the extent to whether Method 3 generically
reproduces the correct result of Method 2, one needs to consider a
wide variety of models and compare the results.  Since the
effective potential only depends on the Skyrme profile function we
have considered a variety of profile functions  (which we
constructed on an {\it ad hoc} basis entirely for the purpose of
testing the validity of Method 3). The curvature at zero recoil of
the Isgur-Wise function was calculated for these effective
potentials using the three methods detailed above.  The effective
potentials that were considered are shown in Fig.~\ref{fig:effpot}
while the results of the calculations of $\rho$ are presented in
Table 1. One can clearly see that with all effective potentials
considered, the curvature at zero recoil calculated with harmonic
wave functions is different from the calculations performed with
the other two methods. However, for all the potentials considered,
the latter two methods provide similar results. Although not
depicted in Table 1, when an effective potential with a global
minimum away from the origin is considered, Methods 2 and 3 give
different results. This can be attributed to the wave function
calculated with the complete effective potential being peaked away
from the origin, where Method 3 assumes that the wave function is
still peaked at the origin.

\begin{table}[htb]
\begin{tabular}{|c|c|c|c|}
\hline Potential& Method 1 & Method 2 & Method 3\\
\hline a (Original) & -53.8 & -121 & -111\\
 b& -83.3& -175 &-171\\
c& -109& -216 & -215\\
d&  -101&-142 & -139\\
e&  -132&-225 & -231\\
\hline
\end{tabular}
\caption{The curvature at zero recoil, $\rho$, of the Isgur-Wise
function calculated in three different manners in units of
$10^{-6} \text{MeV}^{-3/2}$. The first line is the original
profile function we considered.  All calculations were performed
with the coupling constant, $g'$, chosen to bind $\Lambda_b$ with
the appropriate binding energy.}
\end{table}

The fact that Method 3 works so well, even when the system is
quite anharmonic, appears to have some important consequences.  In
the first place, it means the prediction of $\rho$ from the
excited state energy (using the harmonic approximation) may be
expected to hold reasonably well and, hence, one has some real
predictive power even if the system is rather anharmonic.  The
converse of this appears to be that the degree to which $\rho$ is
accurately predicted via Method 3 is a poor test of the degree to
which the system is harmonic.

In fact, the situation is a bit more subtle than this.
Qualitatively it is clear what is happening: the anharmonic nature
of the potential lowers the excitation energy compared to that of
a harmonic potential with the same curvature at the minimum.
Fitting this excitation with a harmonic oscillator means that the
fitted oscillator will have a smaller curvature ({\it i.e.},
spring constant)  than the actual spring constant at the minimum.
This has the effect of spreading out the wave function compared to
the harmonic approximation based on the true curvature which in
turn means a larger value of $\rho$; this acts to simulate the
true wave function which is also wider than the naive harmonic
result. Thus, generically, the {\it sign} of the effect of
anharmonicity on the excitation energy and $\rho$ helps explain
the viability of Method 3. The degree to which the method works
quantitatively may still seem remarkable, however. The
quantitative success is at least partially understandable
analytically.  It is straightforward to calculate the leading
order effect of the anharmonicity on both the excitation energy of
the lowest excited state \cite{ck2}.  As it happens, the shift in
the excitation energy {\it exactly} compensates the shift in
$\rho$ to leading order in the anharmonicity, and the according
inaccuracies due to using Method 3 only appear at next-to-next-to
leading order.  Thus, the system can be rather anharmonic and
Method 3 can remain reasonably accurate.  It is nevertheless
remarkable how well Method 3 appears to work since the wave
functions appear to be qualitatively quite different from the
harmonic ones of Method 1.

In conclusion, we have constructed the effective potential for the
binding of the heavy (anti-)meson and nucleon to form regular and
exotic heavy baryons for variations of the Skyrme model. We have
demonstrated that this effective potential gives  excitation
energies and collective wave functions which are qualitatively
different from those obtained with a harmonic oscillator
approximation to the potential: the wave functions with realistic
particle masses are not concentrated near the potential minimum,
as expected from the large $N_c$ and heavy quark limits. This
indicates that the masses are not heavy enough to have heavy
baryons exhibit the properties of these limits. Calculations with
the full effective potential showed that heavy pentaquark states
are possible in this class of model but the existence of bound
pentaquarks depends sensitively on the details of the model
studies. We also showed that despite strong anharmonicities the
description of the Isgur-Wise function derived from the harmonic
approximation works remarkably well provided that the effective
harmonic coupling derived from the first excitation energy is
used.

{\it Acknowledgments.}  T.D.C.\ and P.M.H.\ were supported by the
D.O.E.\ through grant DE-FGO2-93ER-40762.

\end{document}